\documentclass[12pt]{article}
\usepackage[a4paper, total={7in, 10in}]{geometry}
\usepackage[parfill]{parskip}
\usepackage{physics, tensor, float, subcaption}
\usepackage{graphicx}
\graphicspath{ {Plots/} }
\usepackage{jhep-mod}
\usepackage{bm}
\usepackage{soul}
\usepackage{amssymb,amsmath,amsthm}
\usepackage{mathrsfs}
\usepackage[utf8]{inputenc}
\usepackage{enumerate}
\usepackage{bigints}
\usepackage{xcolor}
\usepackage{appendix}
\usepackage{graphicx}
\usepackage{float}
\usepackage{tikz}
\usepackage{setspace}
\usepackage{cancel}
\usepackage{doi}
\definecolor{purple}{rgb}{1,0,1}
\definecolor{lime}{HTML}{A6CE39} 

\definecolor{lime}{HTML}{A6CE39}
\newcommand{\orcidicon}{%
	\begin{tikzpicture}
	\draw[lime, fill=lime] (0,0) 
		circle [radius=0.16] 
		node[white] {{\fontfamily{qag}\selectfont \tiny ID}};
	\draw[white, fill=white] (-0.0625,0.095) 
		circle [radius=0.007];
	\end{tikzpicture}
	\hspace{-5mm}
}

\newcommand\orcidAlex{{\href{https://orcid.org/0000-0002-1763-3563}{\orcidicon}}}
\newcommand\orcidMatt{{\href{https://orcid.org/0000-0003-1088-6485}{\orcidicon}}}

\newcommand{\e}{\mathrm{e}}

\renewcommand{\O}{\mathcal{O}}

\begin{document}

\title{
The eye of the storm: \\
A regular Kerr black hole
}
\author{
\Large
Alex Simpson\!\orcidAlex\! and Matt Visser\!\orcidMatt\!
}
\affiliation{School of Mathematics and Statistics, Victoria University of Wellington, \\
\null\qquad PO Box 600, Wellington 6140, New Zealand.}
\emailAdd{alex.simpson@sms.vuw.ac.nz}
\emailAdd{matt.visser@sms.vuw.ac.nz}

\abstract{
\vspace{1em}

We analyse in some detail a highly tractable non-singular modification of the Kerr geometry, dubbed the ``eye of the storm'' --- a rotating regular black hole with an asymptotically Minkowski core. This is achieved by ``exponentially suppressing'' the mass parameter in the Kerr spacetime: $m\to m \; \e^{-\ell/r}$. The single suppression parameter $\ell$ quantifies the deviation from the usual Kerr spacetime. Some of the classical energy conditions are globally satisfied, whilst certain choices for $\ell$ force any energy-condition-violating physics into the deep core. The geometry possesses the full ``Killing tower'' of principal tensor, Killing--Yano tensor, and nontrivial Killing tensor, with associated Carter constant; hence the Hamilton--Jacobi equations are separable, and the geodesics integrable. Both the Klein--Gordon equation and Maxwell's equations are also separable on this candidate spacetime. The tightly controlled deviation from Kerr renders the physics extraordinarily tractable when compared with analogous candidates in the literature. This spacetime will be amenable to straightforward extraction of astrophysical observables falsifiable/ verifiable by the experimental community. 

\bigskip

\bigskip
\noindent
{\sc Date:} 24 November 2021; 3 December 2021; \LaTeX-ed \today

\bigskip
\noindent{\sc Keywords}:
Kerr spacetime, rotating regular black hole, Minkowski core, black hole mimic, axisymmetry, LIGO/Virgo, LISA, Killing tower, energy conditions.

\bigskip
\noindent{\sc PhySH:} 
Gravitation
}

\maketitle
\def\tr{{\mathrm{tr}}}
\def\diag{{\mathrm{diag}}}
\def\cof{{\mathrm{cof}}}
\def\pdet{{\mathrm{pdet}}}
\def\d{{\mathrm{d}}}
\parindent0pt
\parskip7pt
\def\Kerr{{\scriptscriptstyle{\mathrm{Kerr}}}}
\def\eos{{\scriptscriptstyle{\mathrm{eos}}}}
\section{Introduction}

Classical curvature singularities, resulting from gravitational collapse, typically occur at distance scales where there is no longer any empirical reason to believe that general relativity is applicable. Consequently, there are at least two routes available to the aspiring relativist:
\begin{itemize}
    \item Try to build a full-fledged and phenomenologically verifiable theory of quantum gravity from scratch (hard).
    \item Purely classically, excise curvature singularities from GR in astrophysically appropriate regimes, and extract associated astrophysical observables which are at least in principle falsifiable/verifiable by the experimental communities now operating in observational and gravitational wave astronomy (nontrivial, but comparably straightforward).
\end{itemize}
Recent experimental successes have greatly enhanced humanity's ability to probe theoretical predictions concerning astrophysical objects. These include the direct observation of gravitational waves emanating from an astrophysical source in the LIGO/Virgo merger events~\cite{LIGO1, LIGO2}, as well as the pioneering image of the black hole in M$87$ by the Event Horizon Telescope (EHT)~\cite{EHT1, EHT2, EHT3, EHT4, EHT5, EHT6}. 
Combined with the planned experimental capabilities of the upcoming LISA project~\cite{LISA}, and planned next-generation ground-based observatories~\cite{science-book}, it is becoming increasingly desirable for theoreticians to compile results for mathematically tractable, curvature-singularity-free candidate spacetimes that speak to the advances made in the experimental community. It is a very real hope that phenomenological evidence will then be obtained which enables us to delineate between candidate spacetimes based on their astrophysical signature. Not only will this process paint a more accurate picture concerning curvature singularities, but it will also give the theoretical community experimentally-informed clues as to which specific modifications to the Einstein equations, or indeed to theoretical physics in general, might be necessary in constructing a ``theory of everything''.

Exploration regarding the extraction of astrophysical observables for
nonsingular candidate geometries has been performed in a vast array of contexts~\cite{Eiroa:2010, Flachi:2012, Abdujabbarov:2016, Carballo-Rubio:2018, Carballo-Rubio:2019a, Carballo-Rubio:2019b, Dai:2019, Cramer:1994, Simonetti:2020, Berry:2020, Carballo-Rubio:2021a,  Carballo-Rubio:2021b, Bronnikov:2021, Churilova:2021, Bambi:2021, Simpson:2021biv}. One could propose a list of further constraints on such geometries which would render them as appropriate as possible for the experimental community, and streamline the discourse between theory and experiment. For instance, an ``idealised'' candidate spacetime could be asked to satisfy at least the following constraints: \enlargethispage{40pt}
\begin{itemize}
    \item Astrophysical sources rotate --- impose axisymmetry.
    \item Impose asymptotic flatness at spatial infinity~\cite{PGLT1}.
    \item 
    The Hamilton--Jacobi equations should be separable ---  the geodesic equations should be at least numerically integrable to enable direct comparison with experimental data. (A sufficient condition for this in axisymmetry is the existence of a nontrivial Killing 2-tensor $K_{\mu\nu}$.)
    \item 
    Impose separability of both Maxwell's equations and the equations governing the spin two polar and axial modes on the background spacetime --- this allows for the ``standard'' numerical techniques to be applied in analysing the quasi-normal modes of spin-one electromagnetic and spin-two gravitational perturbations on the background spacetime, which are fundamental in analysing the ringdown phase of binary mergers. (A good mathematical precursor for this constraint is the separability of the Klein--Gordon equation.)
    \item 
    Impose a high degree of mathematical tractability. The complex process beginning with the inception of a candidate geometry, and finishing with a result able to be directly compared with experimental measurement involves many nontrivial steps --- candidate spacetimes amenable to highly tractable mathematical analysis will yield their astrophysical observables with \emph{far} more ease. 
    \item 
    Constrain the amount of exotic matter and demand satisfaction of the relevant classical energy conditions outside horizons --- empirical evidence suggests any violation of the classical energy conditions should occur at a quantum scale, and (apart from the violations of the SEC due to a positive cosmological constant) we have not observed exotic matter in an astrophysical context~\cite{Visser:epoch,Visser:epoch-prd,Visser:cosmo99}. 
\end{itemize}
The above list serves only as a rough guideline. Naturally, there are many other constraints that are likely to be desirable --- forbid closed timelike curves, for instance, or impose separability of the Dirac equation, \emph{etc.} However the above list speaks \emph{directly} to the current observational and experimental community. Finding appropriate geometries which satisfy \emph{all} of these constraints is highly nontrivial, and generally speaking the best one can do is to use them as goalposts when constructing a candidate spacetime.

A subset of the nonsingular geometries of interest are the so-called ``regular black holes'' (RBHs). By regular, one means in the sense of Bardeen~\cite{Bardeen:1968}, with regularity achieved \emph{via} enforcing global finiteness on orthonormal curvature tensor components and Riemann curvature invariants. In both spherical symmetry and axisymmetry, RBHs have a well-established lineage both in the historical and recent literature~\cite{Bardeen:1968, Ayon-Beato:2000, Hayward:2005, Bronnikov:2005, Bambi:2013, 
Li:2013, Jusufi:2017, Jusufi:2019, Jusufi:2020, Abdujabbarov:2016, 
Carballo-Rubio:2018, Carballo-Rubio:2019a, Carballo-Rubio:2019b,
Carballo-Rubio:2021a, Carballo-Rubio:2021b,
Herdeiro:2016, Frolov:2014, Ghosh:2014, Neves:2014, Toshmatov:2014, quadratic, Fan:2016, Toshmatov:2017a, Toshmatov:2017b, Simpson:2018, Simpson:2019, Lobo:2020a, Simpson:2020, Brahma:2020, Mazza:2021, Franzin:2021}.

Herein we shall explore a rotating RBH with an asymptotically Minkowski core. This geometry was in fact first proposed by Ghosh in reference~\citenum{Ghosh:2014}; we discovered it independently by following a set of carefully chosen metric construction criteria which will be explored in \S~\ref{ansatz}. Consequently, some results are repeated, though with rather different representations and emphasis. Numerous new and important results for this geometry are also presented. Set in stationary axisymmetry, this spacetime is a tightly controlled deviation from standard Kerr, is amenable to highly tractable mathematical analysis, and possesses the full ``Killing tower''~\cite{Frolov:2017} of principal tensor, Killing--Yano tensor, and nontrivial Killing tensor. This induces an associated Carter constant~\cite{Carter:1968a,Carter:1968b}, giving a fourth constant of the motion and rendering the geodesic equations of motion for test particles in principle integrable (\emph{i.e.}, imposing separability of the Hamilton--Jacobi equation). We shall see that any energy-condition-violating physics is able to be pushed  into the deep core, at a distance scale where GR is no longer empirically justified. Both the Klein--Gordon equation and Maxwell's equations are separable on the background spacetime, enabling quasi-normal modes analysis for spin zero and spin one perturbations \emph{via} the `standard' techniques.

With reference to the above list of proposed constraints, comparison with the existing literature on rotating RBHs reveals that this geometry is very close to ``experimentally ideal''.\footnote{It is also a good idea to bear in mind what \emph{cannot} be done: For instance the spatial slices of the Kerr spacetime cannot be put in conformally flat form~\cite{Kroon}, nor can the 3-metric  even be globally diagonalized~\cite{Darboux}. } Before segueing into the analysis of this specific candidate geometry, it is worth exploring the choices made in constructing the metric.

\section{Metric ansatz}\label{ansatz}

To set the stage, we shall first discuss static spherical symmetry for exposition, before migrating the discussion to the more astrophysically appropriate realm of stationary axisymmetry.

\subsection{Spherical symmetry}

Recall one can always put any static, spherically symmetric line element into the standard form~\cite{MTW, Wald, Lorentzianwormholes}
\begin{equation}
    \d s^2 = -\e^{-2\Phi(r)}\left(1-\frac{2m(r)}{r}\right)\d t^2 + \frac{\d r^2}{1-\frac{2m(r)}{r}} + r^2\d\Omega^2_2 \ .
\end{equation}
Further specialising to $\Phi(r)=0$ spacetimes leaves one with the 1-function class of geometries characterised by~\cite{Jacobson:2007}:
\begin{equation}\label{Sch+1}
    \d s^2 = -\left(1-\frac{2m(r)}{r}\right)\d t^2 + \frac{\d r^2}{1-\frac{2m(r)}{r}} + r^2\d\Omega^2_2 \ .
\end{equation}
Specialising to $m(r)=m$ yields the Schwarzschild solution in standard curvature coordinates. Consequently, one can think of equation~(\ref{Sch+1}) as the class of $1$-function modified Schwarzschild geometries. In the discourse that follows, it will be occasionally useful to think from this more general perspective.
Within this 1-function class of geometries, with an eye towards RBHs specifically, one has:\\
{\bf Bardeen \cite{Bardeen:1968}:} 
\begin{equation}
m(r) = {\frac{mr^3}{(r^2+\ell^2)^{3/2}}} \ ; \qquad 
\rho(r)={\frac{m \ell^2}{{\frac{4\pi}{3}} (r^2+\ell^2)^{5/2}}} \ .
\end{equation}
This implies an asymptotically de Sitter core with 
\begin{equation}
\rho(0) ={\frac{m}{{\frac{4\pi}{3}} \ell^3}} \ .
\end{equation}
So the central density depends on asymptotic mass. \\
{\bf Hayward \cite{Hayward:2005}:}
\begin{equation}
m(r) = {\frac{m r^3}{r^3+ 2 m \ell^2}} \ ; \qquad 
\rho(r)={\frac{m^2\ell^2}{{\frac{2\pi}{3}} (r^3+2m\ell^2)^{3}}} \ .
\end{equation}
This implies an asymptotically de Sitter core with 
\begin{equation}
\rho(0) ={\frac{1}{{\frac{8\pi}{3}} \ell^2}} \ .
\end{equation}
So the central density is independent of asymptotic mass.\\
{\bf Asymptotically Minkowski core \cite{Simpson:2020}:}
\begin{equation}
m(r) = m\,\e^{-\ell/r} \ ; \qquad \rho(r) = \frac{\ell m\,\e^{-\ell/r}}{4\pi r^4} \ .
\end{equation}
This implies an asymptotically Minkowski core with
\begin{equation}
\rho(0) = 0 \ .
\end{equation}
So the central density is zero.

The inspiration for the construction of the candidate spacetime analysed herein comes directly from the regular black hole with asymptotically Minkowski core,  presented and analysed in reference~\citenum{Simpson:2020}. The explicit line element is given by:
\begin{equation}\label{AM}
    \d s^2 = -\left(1-\frac{2m\,\e^{-\ell/r}}{r}\right)\,\d t^2 + \frac{\d r^2}{1-\frac{2m\,\e^{-\ell/r}}{r}} + r^2\,\d\Omega^2_2 \ .
\end{equation}
In the $\ell\rightarrow0$ limit, Schwarzschild spacetime in the usual curvature coordinates is recovered precisely. Consequently the ``supression parameter'' $\ell$ can be viewed as quantifying the deviation from Schwarzschild spacetime. When viewed as a modification of Schwarzschild, one performs the following ``regularisation'' procedure:
\begin{itemize}
    \item Make the modification $m\rightarrow m(r)=m\,\e^{-\ell/r}$.
\end{itemize}
It should be emphasised that this is \emph{not} a coordinate transformation.

Due to the severe mathematical discontinuity of the function $\e^{-\ell/r}$ at coordinate location $r=0$, the line element equation~(\ref{AM}) is not $C^{\omega}$ at $r=0$ (it is in fact not even $C^0$). This implies that the region $r<0$ is grossly unphysical for this candidate spacetime. In and of itself, this is not a problem \emph{per se}, and physical analysis is valid for $r\geq0$. However, this raises the question: What are the most prudent mathematical choices one can make when attempting to ``regularise'' a candidate black hole \emph{via} exponential suppression? In the regime of static spherical symmetry, here are two other examples which are worth brief discussion.
\paragraph{Example:} Consider
\begin{equation}
\d s^2 = -\left(1-\frac{2m(r)}{r}\right)\,\d t^2 + \frac{\d r^2}{1-\frac{2m(r)}{r}} + r^2 \; \d \Omega^2_2 \ ;
\qquad
m(r) = m\,\exp\left(-\frac{\ell^2}{r^2}\right) \ .
\end{equation}
(For related ideas, see for instance~\cite{quadratic}.)
Purely mathematically, $\exp\left(-{\ell^2}/{r^2}\right)$ is real analytic only for $r\neq0$, however it is $C^{+\infty}$ for all $r$. Superficially then, this example looks more general than that of equation~(\ref{AM}), due to the fact one can now extend the analysis to $r<0$. A deeper look reveals that this is not particularly useful. On each spatial slice $r=0$ is still a point, and in spacetime $r=0$ is a timelike curve. Consequently, one has two universes, corresponding to $r\geq0$ and $r\leq0$, with each being a copy of the geometry characterised by equation~(\ref{AM}), connected at the single point $r=0$. One may not traverse through this point, and the ``other'' universe is physically irrelevant.
\paragraph{Example:} Consider instead
\begin{equation}\label{Ex2}
\d s^2 = -\left(1-\frac{2m(r)}{r}\right)\,\d t^2 + \frac{\d r^2}{1-\frac{2m(r)}{r}} + (r^2+\ell^2) \; \d \Omega^2_2 \ ;
\qquad
m(r) = m\,\exp\left(-\frac{\ell^2}{r^2}\right) \ .
\end{equation}
Because we have modified the angular part of the metric, this is now intrinsically more general (from a physical perspective). Specifically, on any spatial slice $r=0$ now corresponds to a $2$-sphere of finite area $4\pi\ell^2$, and in spacetime $r=0$ is in fact a timelike hypertube (\emph{i.e.} a traversable wormhole throat). Now the two universes corresponding to $r\leq0$ and $r\geq0$ are connected at the traversable throat $r=0$, and a would-be timelike traveller may propagate between them. The qualitative causal structure has both an outer and an inner horizon, with a timelike traversable hypersurface in the deep core at $r=0$; this is qualitatively the same as for certain specialisations explored in references~\citenum{Mazza:2021, Franzin:2021}.

\subsection{Kerr-like rotating spacetimes}

Instead, herein we are interested in investigating a rotating version of the regular black hole with asymptotically Minkowski core. This is better-motivated from an astrophysical standpoint, and hence more likely to speak to the relevant parties currently operating in observational and gravitational wave astronomy. Migrating the discourse to stationary axisymmetry, we begin with the Kerr spacetime in standard Boyer--Lindquist (BL) coordinates:
\begin{equation}\label{Kerr}
    \d s^2 = - \frac{\Delta_{\Kerr}}{\Sigma}(\d t-a\sin^2\theta\,\d\phi)^2 
    + \frac{\sin^2\theta}{\Sigma}[(r^2+a^2)\,\d\phi-a\,\d t]^2
    +\frac{\Sigma}{\Delta_{\Kerr}}\,\d r^2 + \Sigma\,\d\theta^2
     \ ,
\end{equation}
where as usual
\begin{equation}
    \Sigma = r^2+a^2\cos^2\theta \ , \qquad \Delta_{\Kerr}=r^2+a^2-2mr \ .
\end{equation}
The inverse metric can be written as:
\begin{equation}
g_\Kerr^{\mu\nu} \partial_{\mu} \partial_{\nu} = \frac{1}{\Sigma}\left\{ 
-{\left[(r^2+a^2)\,\partial_t  +a\, \partial_\phi \right]^2\over\Delta_\Kerr}
 +{(\partial_\phi + a\sin^2\theta\, \partial_t)^2 \over\sin^2\theta}
+\Delta_\Kerr \partial_r^2 + \partial_\theta^2\right\} \ .
\end{equation}
In BL coordinates, the ring singularity present in Kerr spacetime is located at $r=0$. We now attempt to ``regularise'' Kerr spacetime. Inspired by the aforementioned procedure for the RBH with asymptotically Minkowski core, we leave the object $\d r$ in the metric undisturbed, and make a modification $m\rightarrow m(r)$. 

Prosaically, this class of geometries can be viewed as a ``$1$-function off-shell'' extension of the Kerr geometry; this is a specialisation of Carter's ``$2$-function off-shell'' extension to Kerr~\cite{Frolov:2017, Carter:1968a, Carter:1968b}. From reference~\citenum{Frolov:2017},  we already know that geometries within this ``$1$-function off-shell'' extension to Kerr must possess a nontrivial Killing tensor $K_{\mu\nu}$. This implies the existence of an associated Carter constant, and hence separability of the Hamilton--Jacobi equations (and, in principle, integrable geodesics). This is yet another motivation for exploring this line of inquiry. In \S~{\ref{Killingtower}} we will explicitly verify that in general the ``$1$-function off-shell'' extension to Kerr always in fact possesses the full ``Killing tower'' of Killing tensor, Killing--Yano tensor, and principal tensor~\cite{Frolov:2017}.

With the ``$1$-function off-shell'' extension to Kerr in hand, another potential approach might be to instead make the modification $m\rightarrow m(r,\theta)$. One \emph{may} intuit from the fact that the slices of axisymmetry are $\theta$-dependent that any exponential ``supression mechanism'' also ought to have a $\theta$-dependence; $\theta$-dependent modifications to certain mass functions in axisymmetry have been discussed in~\cite{Eichhorn:2021a, Eichhorn:2021b}. However, in Kerr spacetime there are also geometric features of qualitative importance which are $\theta$-\emph{independent}, such as the horizon locations. Imposing this $\theta$-dependence also loses the guarantee that one can put the metric into the form of Carter's ``$2$-function off-shell'' extension to Kerr; one \emph{may} lose the existence of a nontrivial Killing tensor $K_{\mu\nu}$ (and of course the associated ``Killing tower'')~\cite{Frolov:2017}. Imposing this $\theta$-dependence also has severe implications on mathematical tractability.

For both approaches considered, fixing the most desirable $m(r)$ or $m(r,\theta)$ such that the candidate geometry is both \emph{regular} and \emph{tractable} is nontrivial, and all of the following examples are worth brief discussion.
\paragraph{Example:} Consider ``$1$-function off-shell'' Kerr (in BL coordinates) with
\begin{equation}
m(r) = m\,\exp\left(-\frac{\ell}{r}\right) \ .
\end{equation}
Mathematically, one has the discontinuity at $r=0$, and hence the region $r<0$ is omitted from the analysis. In terms of Cartesian coordinates $r_{naive}^2 = x^2+y^2+z^2$ one has
\begin{equation}
r_{naive}^2 = r^2 + a^2 - {a^2 z^2\over r^2} \ ; \qquad \cos\theta= z/r \ .
\end{equation}
Then
\begin{equation}
r_{naive}^2 = r^2 + a^2 \sin^2\theta \ ,
\end{equation}
while
\begin{equation}
\cos\theta_{naive}= {z\over r_{naive}} = {z\over r} \;{r\over r_{naive}} = 
\frac{\cos\theta}{\sqrt{ 1 + a^2 \sin^2\theta/r^2}}
=
\frac{r\; \cos\theta}{\sqrt{ r^2 + a^2 \sin^2\theta}} \ .
\end{equation}
So exponential suppression in the BL coordinate $r$ (as $r\to 0^+$) suppresses the mass function $m(r)$ in the entire Cartesian disk $r_{naive} \leq a$, where $\cos\theta_{naive}=0$. This ought to render the spacetime curvature-regular (and indeed does so; this will be demonstrated shortly). This specific example is particularly useful in that the ``other universes'' in the maximal analytic extension of the usual version of Kerr are removed from the analysis due to the restriction $r\geq0$. Consequently the maximal analytic extension of this regularized spacetime will be trivial --- there will be no concerns arising from closed timelike curves in this candidate geometry.\footnote{It is perhaps worthwhile to note that even for standard Kerr spacetime, the closed timelike curves can arise only by dodging into the ``other'' universe ($r<0$). This is most obvious in Doran coordinates where, since $g^{tt}=-1$, the entire $r>0$ region is manifestly stably causal~\cite{Doran, brief-introduction, Kerr-book, unit-lapse}. }

\paragraph{Example:} Consider instead ``$1$-function off-shell'' Kerr (in BL coordinates) with
\begin{equation}
m(r) = m\,\exp\left(-\frac{\ell^2}{r^2}\right) \ .
\end{equation}
(See for instance~\cite{quadratic}.)
Purely mathematically, one may now also consider $r<0$ given the metric is now $C^{+\infty}$ at $r=0$. In terms of Cartesian coordinates $r_{naive}^2 = x^2+y^2+z^2$ one still has
\begin{equation}
r_{naive}^2 = r^2 + a^2 \sin^2\theta \ ;
\end{equation}
\begin{equation}
\cos\theta_{naive}= \frac{r\; \cos\theta}{\sqrt{ r^2 + a^2 \sin^2\theta}} \ .
\end{equation}
So (quadratic) exponential suppression in the BL coordinate $r$ (as $r\to 0^+$) suppresses the mass function $m(r)$ in the entire Cartesian disk $r_{naive} \leq a$, where $\cos\theta_{naive}=0$.\\
 However, physically there is now no point in continuing the $r$ coordinate to $r<0$. In the absence of the ring singularity at $r_{naive}=a$, there is nothing to generate an angle deficit or angle surfeit; the ring at $r_{naive}=a$ is utterly ordinary. Consequently, exploring $r\leq0$ is physically identical to exploring $r\geq0$. Notably, the curvature quantities and general analysis for this example are less tractable than for the example based on $\exp(-\ell/r)$.

%
\paragraph{Example:} Consider modified Kerr (in BL coordinates) with the somewhat messier $\theta$-dependent mass function
\begin{equation}
m(r,\theta) = m\,\exp\left(-\frac{\ell}{\sqrt{r^2+a^2\cos\theta^2}}\right) = m\,\exp\left(-\frac{\ell}{\sqrt{\Sigma}}\right) \ .
\end{equation}
Now $\sqrt{r^2+a^2\cos\theta^2}=0$ requires both $r=0$ and $\theta=\pi/2$. There will be the discontinuity at $r=0$ in the equatorial plane, where $m(r,\theta)\rightarrow m\,\exp\left(-\frac{\ell}{r}\right)$; one must omit $r<0$ from the analysis. Furthermore, \emph{via} the standard results
\begin{equation}
r_{naive}^2 = r^2 + a^2 \sin^2\theta \ ,
\end{equation}
and
\begin{equation}
\cos\theta_{naive}= \frac{r\; \cos\theta}{\sqrt{ r^2 + a^2 \sin^2\theta}} \ ,
\end{equation}
this implies that both 
$r_{naive}=a$ and $\theta_{naive}=\pi/2$. So exponential suppression in $\sqrt{r^2+a^2\cos^2\theta}$ suppresses the mass function $m(r,\theta)$ only at the \emph{edge} of the Cartesian {disk} $r_{naive} = a$, where $\cos\theta_{naive}=0$. The geometry is now not flat on the interior of the disk $r_{naive} < a$, with $\cos\theta_{naive}=0$. Supplementary to this, imposing the $\theta$-dependence in this specific manner has \emph{severe} implications on the tractability of the analysis.
\enlargethispage{40pt}
\paragraph{Example:} Consider instead modified Kerr (in BL coordinates) with the messier $\theta$-dependent mass function
\begin{equation}
    m(r,\theta) = m\,\exp\left(-\frac{\ell^2}{r^2+a^2\cos^2\theta}\right) = m\,\exp\left(-\frac{\ell^2}{\Sigma}\right) \ .
\end{equation}
Note that one may now explore, purely mathematically, $r<0$. By the same logic as for the previous example, exponential suppression in $r^2+a^2\cos^2\theta$ suppresses the mass function $m(r,\theta)$ only at the \emph{edge} of the Cartesian disk $r_{naive}=a$, where $\cos\theta_{naive}=0$. The geometry is not flat on the interior of this disk. There is no ring singularity at $r_{naive}=a$, and so nothing to generate an angle deficit or angle surfeit; the ring at $r_{naive}=a$ is utterly ordinary. Consequently, even though one may mathematically explore $r<0$, there is no physical reason to do so, by the same logic as for previous examples. Furthermore, imposing the $\theta$-dependence in this manner \emph{severely} affects mathematical tractability.

Ultimately, deciding which candidate geometry is preferable for analysis is nontrivial. Exploring these examples has left us with the following conclusions:
\begin{itemize}
    \item There is no physical point to forcing the analysis to be amenable to analytic extension to $r<0$ in this specific manner, and doing so has consequences concerning mathematical tractability;
    \item There may or may not be a physical point to forcing the suppression mechanism to have a $\theta$-dependence, however doing so has \emph{severe} implications on mathematical tractability, and also does not render the central disk Minkowski.
\end{itemize}
%

\subsection{The eye of the storm}

Consequently, we advocate for the most mathematically tractable of the aforementioned examples in axisymmetry; this is the example $m(r)=m \exp(-\ell/r)$. This results in the following specific and fully explicit metric, for now labelled the ``eye of the storm'' (eos) spacetime:
\begin{equation}\label{E:eos}
    \d s^2 = \frac{\Sigma}{\Delta_{\eos}}\,\d r^2 + \Sigma\,\d\theta^2 - \frac{\Delta_{\eos}}{\Sigma}(\d t-a\sin^2\theta\,\d\phi)^2 + \frac{\sin^2\theta}{\Sigma}[(r^2+a^2)\,\d\phi-a\,\d t]^2 \ ,
\end{equation}
where
\begin{equation}
    \Sigma = r^2+a^2\cos^2\theta \ , \qquad \Delta_{\eos}=r^2+a^2-2mr\,\e^{-\ell/r} \ .
\end{equation}
The inverse metric can be written as:
\begin{equation}
g_\eos^{\mu\nu} \partial_{\mu} \partial_{\nu} = \frac{1}{\Sigma}\left\{ 
-{\left[(r^2+a^2)\,\partial_t  +a\, \partial_\phi \right]^2\over\Delta_\eos}
 +{(\partial_\phi + a\sin^2\theta\, \partial_t)^2 \over\sin^2\theta}
+\Delta_\eos \partial_r^2 + \partial_\theta^2\right\} \ .
\end{equation}
We note again that this is the same geometry as presented by Ghosh in reference~\citenum{Ghosh:2014}, now with a considerably more detailed physical justification as to why it is of interest. In the limit as $r\rightarrow+\infty$, asymptotic flatness is preserved. In the limit as $\ell\rightarrow0$, one returns the standard Kerr spacetime in BL coordinates. As such, we enforce $\ell>0$ for nontrivial analysis, and the supression parameter $\ell$ can be viewed as quantifying the deviation from Kerr spacetime. In the limit as $a\rightarrow0$, one recovers equation~(\ref{AM}) precisely. In comparison with standard Kerr in BL coordinates, the domains for the temporal and angular coordinates are unaffected. However the discontinuity at $r=0$ restricts the domain for the $r$ coordinate to $r\geq0$. This removes concerns involving closed timelike curves which are present in the ``usual'' discourse surrounding maximally extended Kerr. Crucially, as we shall shortly observe, the ring singularity is excised; replaced instead by a region of spacetime which is asymptotically Minkowski. This renders the geometry globally nonsingular, and we have a tractable model for a regular black hole with rotation.

From the form of the line element as in equation~(\ref{E:eos}), ordering the coordinates as $(t,r,\theta,\phi)$, it is straightforward to read off a convenient covariant tetrad (co-tetrad) which is a solution of $g_{\mu\nu}=\eta_{\hat{\mu}\hat{\nu}}\;e^{\hat{\mu}}{}_{\mu}\; e^{\hat{\nu}}{}_{\nu}$ (it should be noted this co-tetrad is not unique):
\begin{eqnarray}
    e^{\hat{t}}{}_{\mu} &=& \sqrt{\frac{\Delta_{\text{eos}}}{\Sigma}}\left(-1;0,0,a\sin^2\theta\right) \ ; \qquad \qquad e^{\hat{r}}{}_{\mu} = \sqrt{\frac{\Sigma}{\Delta_{\text{eos}}}}\left(0;1,0,0\right) \ ; \nonumber \\
    && \nonumber \\
    e^{\hat{\theta}}{}_{\mu} &=& \sqrt{\Sigma}\left(0;0,1,0\right) \ ; \qquad \qquad e^{\hat{\phi}}{}_{\mu} = \frac{\sin\theta}{\sqrt{\Sigma}}\left(-a;0,0,r^2+a^2\right) \ .
\end{eqnarray}
This co-tetrad uniquely defines a contra-tetrad (contravariant tetrad, or just tetrad) \emph{via} $e_{\hat{\mu}}{}^{\mu}=\eta_{\hat{\mu}\hat{\nu}}\; e^{\hat{\nu}}{}_{\nu}\; g^{\nu\mu}$. 

Explicitly:
\begin{eqnarray}
    e_{\hat{t}}{}^{\mu} &=& -\frac{1}{\sqrt{\Sigma\Delta_{\text{eos}}}}\left(r^2+a^2;0,0,a\right) \ ; \qquad e_{\hat{r}}{}^{\mu} = \sqrt{\frac{\Delta_{\text{eos}}}{\Sigma}}\left(0;1,0,0\right) \ ; \nonumber \\
    && \nonumber \\
    e_{\hat{\theta}}{}^{\mu} &=& \frac{1}{\sqrt{\Sigma}}\left(0;0,1,0\right) \ ; \qquad e_{\hat{\phi}}{}^{\mu} = \frac{1}{\sqrt{\Sigma\sin^2\theta}}\left(a\sin^2\theta;0,0,1\right) \ .
\end{eqnarray}
This tetrad will be employed to convert relevant tensor coordinate components into an orthonormal basis.

Now, we find it useful to define the object
\begin{equation}
    \Xi = \frac{\ell\Sigma}{2r^3} \ .
\end{equation}
This will greatly simplify some of the following analysis. Where convenient for exposition, curvature quantities are displayed in the form: 
\begin{equation}
\hbox{(something dimensionful)} \times \hbox{(something dimensionless)} \ .
\end{equation}
Note that $\Xi$ is dimensionless.

To confirm the assertion that the eye of the storm is curvature-regular, let us analyse the nonzero components of the Riemann curvature tensor with respect to this orthonormal basis.
Fully explicitly, they are given by
\begin{eqnarray}
    R^{\hat{t}\hat{r}}{}_{\hat{t}\hat{r}} &=& \frac{2r^3m\,\e^{-\ell/r}}{\Sigma^3}\left[2\Xi^2-4\Xi+1-3\left(\frac{a}{r}\right)^2\cos^2\theta\right] \ , \nonumber \\
    && \nonumber \\
    -\frac{1}{2}R^{\hat{t}\hat{r}}{}_{\hat{\theta}\hat{\phi}} = -R^{\hat{t}\hat{\theta}}{}_{\hat{r}\hat{\phi}} = R^{\hat{t}\hat{\phi}}{}_{\hat{r}\hat{\theta}} &=& \frac{a\cos\theta r^2m\,\e^{-\ell/r}}{\Sigma^3}\left[2\Xi+\left(\frac{a}{r}\right)^2\cos^2\theta-3\right] \ , \nonumber \\
    && \nonumber \\
    R^{\hat{t}\hat{\theta}}{}_{\hat{t}\hat{\theta}} = R^{\hat{t}\hat{\phi}}{}_{\hat{t}\hat{\phi}} = R^{\hat{r}\hat{\theta}}{}_{\hat{r}\hat{\theta}} = R^{\hat{r}\hat{\phi}}{}_{\hat{r}\hat{\phi}} &=& \frac{r^3m\,\e^{-\ell/r}}{\Sigma^3}\left[2\Xi+3\left(\frac{a}{r}\right)^2\cos^2\theta-1\right] \ , \nonumber \\
    && \nonumber \\
    R^{\hat{\theta}\hat{\phi}}{}_{\hat{\theta}\hat{\phi}} &=& \frac{2r^3m\,\e^{-\ell/r}}{\Sigma^3}\left[1-3\left(\frac{a}{r}\right)^2\cos^2\theta\right] \ .
\end{eqnarray}
All are of the general form
\begin{equation}
    R^{\hat{\alpha}\hat{\beta}}{}_{\hat{\mu}\hat{\nu}} = \frac{m\, \e^{-\ell/r}}{r^n\,\Sigma^3}X(r,\theta; a, \ell) \ ,
\end{equation}
where the object $X(r,\theta;a,\ell)$ is globally well-behaved. The only potentially dangerous behaviour comes from the $r^n\,\Sigma^3$ present in the denominators in the limit as $r\rightarrow0^{+}$. However the exponential dominates; $\lim_{r\rightarrow0^{+}}\e^{-\ell/r}/(r^n\,\Sigma^3)=0$ for all $\theta$. Consequently the ring singularity present at $r=0$ in BL coordinates for Kerr is replaced by a region of spacetime that is asymptotically Minkowski. This is already enough to conclude that the spacetime is globally regular in the sense of Bardeen~\cite{Bardeen:1968}, and is consistent with the findings in reference~\citenum{Ghosh:2014}.

Note that because the spacetime is now stationary rather than static, the Kretschmann scalar need no longer be positive definite~\cite{Lobo:2020a}. 
It is now not sufficient to examine the Kretschmann scalar for regularity; one needs to inspect all the individual orthonormal Riemann components.

\enlargethispage{40pt}
More generally, for the family of ``$1$-function off-shell'' Kerr geometries all nonzero orthonormal components of the Riemann tensor can be represented by
\begin{equation}
    R^{\hat{\alpha}\hat{\beta}}{}_{\hat{\mu}\hat{\nu}} = Z(r,\theta,m(r),m'(r),m''(r);m,a,\ell) \ ,
\end{equation}
for some function $Z$, and the condition for curvature regularity reduces to
\begin{equation}
    m(r) = \mathcal{O}(r^3) \ .
\end{equation}
The condition for an asymptotically Minkowski core reduces to
\begin{equation}
    m(r) = o(r^3) \ .
\end{equation}

\section{Geometric analysis}
\subsection{Curvature invariants}

For the sake of rigour, let us examine the Riemann curvature invariants associated with the candidate geometry.

The Ricci scalar is given by
\begin{equation}
    R = \frac{2\ell^2m\,\e^{-\ell/r}}{\Sigma\,r^3} \ .
\end{equation}
The Ricci contraction $R_{\alpha\beta}R^{\alpha\beta}$ is given by
\begin{equation}
    R_{\alpha\beta}R^{\alpha\beta} = \frac{8\ell^2\left(m\,\e^{-\ell/r}\right)^2}{\Sigma^4}\left(\Xi^2-2\Xi+2\right) \ .
\end{equation}
Note that $\left(\Xi^2-2\Xi+2\right) = 1 + (\Xi-1)^2 \geq 1$ is manifestly positive.

The Kretschmann scalar ($K=R_{\alpha\beta\mu\nu}R^{\alpha\beta\mu\nu}$) is given by 
\begin{eqnarray}
    K &=& \frac{48r^6\left(m\,\e^{-\ell/r}\right)^2}{\Sigma^6}\Bigg\lbrace 1 -15\left(\frac{a}{r}\right)^2\cos^2\theta +15\left(\frac{a}{r}\right)^4\cos^4\theta-\left(\frac{a}{r}\right)^6\cos^6\theta \nonumber \\
    && \nonumber \\
    && +\frac{4}{3}\Xi^4-\frac{16}{3}\Xi^3+8\Xi^2\left[1-\left(\frac{a}{r}\right)^2\cos^2\theta\right]-4\Xi\left[1-6\left(\frac{a}{r}\right)^2\cos^2\theta+\left(\frac{a}{r}\right)^4\cos^4\theta\right]\Bigg\rbrace \ . \nonumber \\
    &&
\end{eqnarray}
Note the presence of both positive definite and negative definite terms, with the negative definite terms depending on even powers of the spin parameter $a$, so that they switch off as the rotation is set to zero. Indeed as $a\to 0$ we have $\Xi \to {\ell\over 2r}$ and so
\begin{eqnarray}
    K_{a\to0} &\to& \frac{48\left(m\,\e^{-\ell/r}\right)^2}{r^6}
    \Bigg\lbrace \frac{4}{3}\Xi^4-\frac{16}{3}\Xi^3
    +8\Xi^2 -4\Xi  +1 \Bigg\rbrace \nonumber \\
    &\to &\frac{48\left(m\,\e^{-\ell/r}\right)^2}{r^6}
    \Bigg\lbrace 
     (1-\Xi)^4 +{\Xi^2\,(\Xi-2)^2\over 3} +{2\over3} \, \Xi^2
    \Bigg\rbrace \ .
\end{eqnarray}
This is now manifestly a positive definite sum of squares, as required.

To evaluate the Weyl contraction note that in this situation the (orthonormal) Weyl tensor has only two algebraically independent components
\begin{eqnarray}
C_{\hat t\hat\phi\hat t \hat\phi} &=& -{1\over2}C_{\hat t\hat r\hat t \hat r} =
C_{\hat t\hat\theta \hat t \hat\theta} = -C_{\hat\phi\hat r\hat \phi\hat r} 
= {1\over2} C_{\hat\phi\hat \theta\hat \phi\hat \theta} 
= -C_{\hat r \hat \theta\hat r \hat \theta} \ ;
\\[5pt]
C_{\hat t\hat \phi\hat r\hat\theta} &=&
 {1\over2} C_{\hat t\hat r\hat\phi\hat\theta} 
 = C_{\hat t\hat \theta\hat\phi\hat r} \ ,
\end{eqnarray}
where explicitly
\begin{eqnarray}
C_{\hat t\hat\phi\hat t \hat\phi} &=& \frac{r^3m\,\e^{-\ell/r}}{3\Sigma^3}\left\lbrace 2\Xi^2-6\Xi+3-9\left(\frac{a}{r}\right)^2\cos^2\theta\right\rbrace \ ;
\\
C_{\hat t\hat r\hat\phi\hat\theta} &=& \frac{r^2m\,\e^{-\ell/r}a\cos\theta}{\Sigma^3}\left\lbrace2\Xi-3+\left(\frac{a}{r}\right)^2\cos^2\theta\right\rbrace \ .
\end{eqnarray}
The Weyl contraction ($C_{\alpha\beta\mu\nu}C^{\alpha\beta\mu\nu}$) 
is given by 
\begin{equation}
    C_{\alpha\beta\mu\nu}C^{\alpha\beta\mu\nu} = 
    48 ([C_{\hat t\hat\phi\hat t \hat\phi}]^2 
    - [C_{\hat t\hat \phi\hat r\hat\theta}]^2) \ .
\end{equation}
Note the presence of both positive definite and negative definite terms, with the negative definite terms depending on even powers of the spin parameter $a$, so that they switch off as the rotation is set to zero.

It is then easy to check that $C_{\alpha\beta\mu\nu}C^{\alpha\beta\mu\nu}
= K + 2 R_{\mu\nu} R^{\mu\nu} - {1\over3} R^2$, as also required.
All of these Riemann curvature invariants are globally well-behaved, remaining finite $\forall\,\,r\in[0,+\infty)$, cementing the fact that the eye of the storm is curvature-regular. In the limit as $\ell\rightarrow0$, the expected limiting behaviour when compared with standard Kerr spacetime is observed. In the limit as $a\rightarrow0$, all Riemann invariants tend to their counterparts for the spherically symmetric candidate geometry analysed in reference~\citenum{Simpson:2020}.

\subsection{Ricci and Einstein tensors}

Both the Ricci and Einstein tensors are diagonal in the orthonormal basis. The Ricci tensor is given by
\begin{equation}
    R^{\hat{\mu}}{}_{\hat{\nu}} = \frac{2\ell m\,\e^{-\ell/r}}{\Sigma^2}\,\text{diag}\left(\Xi-1,\Xi-1,1,1\right) \ ,
\end{equation}
and the Einstein tensor is given by
\begin{eqnarray}
    G^{\hat{\mu}}{}_{\hat{\nu}} = -\frac{2\ell m\,\e^{-\ell/r}}{\Sigma^2}\,\text{diag}\left(1,1,\Xi-1,\Xi-1\right) \ .
\end{eqnarray}
These representations are \emph{highly} tractable when compared with the analogous results for other candidate rotating regular black holes in the literature~\cite{Bambi:2013, Li:2013, Jusufi:2017, Jusufi:2019, Jusufi:2020, Abdujabbarov:2016, Herdeiro:2016,Toshmatov:2014, Toshmatov:2017a, Toshmatov:2017b, Mazza:2021, Franzin:2021}.

\subsection{Causal structure and ergoregion}

Horizon locations are characterised by the roots of $\Delta_{\eos}(r)$, which are also the only coordinate singularities present in the line element equation~(\ref{E:eos}). Since $\Delta_{\eos}(r)$ is real,  while $\Delta_{\eos}(r=0) = a^2>0$ and $\Delta_{\eos}(r\to\infty) = \O(r^2)$, there are either two distinct roots, one double root, or zero roots. Since $\Delta_{\eos}(r) > \Delta_{\Kerr}(r)$ the location of the roots of $\Delta_{\eos}(r)$ is trivially bounded by the location of the roots of $\Delta_{\Kerr}(r)$. Specifically
\begin{equation}
m-\sqrt{m^2-a^2} < r_H^- \leq r_H^+ < m+\sqrt{m^2-a^2} \ .
\end{equation}
In particular, if $m<a$ there certainly are no roots.

Analytically, we cannot explicitly solve for the roots of $\Delta_{\eos}(r)$.
However what we can do, assuming the existence of distinct roots $r_H^\pm$, is to ``reverse engineer'' by solving for 
$m(r_H^+,r_H^-)$ and $a^2(r_H^+,r_H^-)$. 
We note that by definition
\begin{equation}
   (r_H^+)^2  -2 m \, r_H^+ \exp( - \ell/r_H^+) + a^2 =0 \ ;
\end{equation}
\begin{equation}
   (r_H^-)^2  -2 m \, r_H^- \exp( - \ell/r_H^-) + a^2 =0 \ .
\end{equation}
These are two simultaneous equations linear in $m$ and $a^2$. 
We find
\begin{equation}
m(r_H^+,r_H^-) =  {(r_H^+)^2-(r_H^-)^2 \over 
2 ( \e^{-\ell/r_H^+} \; r_H^+ - \e^{-\ell/r_H^-} \; r_H^-)} \ ;
\end{equation}
and
\begin{equation}
a^2 (r_H^+,r_H^-) = {r_H^+ r_H^-  (
\e^{-\ell/r_H^-} \; r_H^+ - \e^{-\ell/r_H^+} \; r_H^-)
\over  \e^{-\ell/r_H^+} \; r_H^+ - \e^{-\ell/r_H^-} \; r_H^-} \ .
\end{equation}
In the degenerate extremal limit $r_H^+\to r_H \leftarrow r_H^-$, using the l'H\^opital rule,  this simplifies to
\begin{equation}
m(r_H) =  {(r_H)^2 \; e^{\ell/r_H} \over 
r_H + \ell} > r_H \ ;
\qquad
\hbox{and}
\qquad
a^2 (r_H) = {r_H^2  (r_H-\ell)\over r_H+\ell} < r_H^2 \ .
\end{equation}
For fixed $\ell$ and $r_H$, setting $a\to a(r_H)$, we have: (1) If $m>m(r_H)$ there will be two distinct roots, one above and one below $r_H$. (2) If $m=m(r_H)$ there is one degenerate root exactly at $r_H$. (3) If $m<m(r_H)$ there are no real roots.

\enlargethispage{20pt}
Given this is the best one can say analytically, and that in this context the parameter $\ell$ is often associated with the Planck scale, we may Taylor series expand about $\ell=0$ for an approximation. 

Let us write
\begin{equation}
    r_{H} = m\,\e^{-\ell/r_{H}^\pm} + S_1 \sqrt{m^2\,\e^{-2\ell/r_{H}}-a^2} \ ,
\end{equation}
where $S_1=\pm 1$.
For small $\ell$, to second-order we find
\begin{equation}
    r_{H} = m + S_{1}\sqrt{m^2-a^2-2m\ell-\mathcal{O}(\ell^2)} \ .
\end{equation}
This has the correct limiting behaviour as $\ell\rightarrow0$. Investigating in more detail, instead of expanding about $\ell=0$ we can instead search for the approximate horizon locations by expanding about the Kerr horizon located at $r=r_{H,\Kerr}=m+S_{1}\sqrt{m^2-a^2}$. 

To second-order this gives
\begin{eqnarray}
    r_{H} &=& m + S_{1}\sqrt{m^2-a^2} \nonumber \\
    && \nonumber \\
    && \ -S_{1}\frac{m\left[2m\sqrt{m^2-a^2}+S_{1}(2m^2-a^2)\right]}{(S_{1}m+\sqrt{m^2-a^2})\left[m\sqrt{m^2-a^2}+S_{1}(m^2-a^2)\right]}\ell + \mathcal{O}\left(\ell^2\right) \nonumber \\
    && \nonumber \\
    &=& r_{H,\Kerr} -\frac{2m\left(r_{H,\Kerr}\right)-a^2}{\left(r_{H,\Kerr}\right)\left[r_{H,\Kerr}-\frac{a^2}{m}\right]}\,\ell + \mathcal{O}(\ell^2) \ .
\end{eqnarray}
Notably, the surface area of each horizon is qualitatively unchanged from Kerr spacetime, given by
\begin{equation}
    A_{H} = 2\pi\int_{0}^{\pi}\sqrt{g_{\theta\theta}\,g_{\phi\phi}}\bigg\vert_{r_{H}}\,\d\theta = 4\pi(r_{H}^2+a^2) \ .
\end{equation}
The ergosurface is characterised by $g_{tt}=0$, \emph{implicitly} given by
\begin{equation}
    r_{\text{erg}}^2+a^2\cos^2\theta = 2mr_{\text{erg}}\,\e^{-\ell/r_{\text{erg}}} \ ,
\end{equation}
and for small $\ell$, is to second-order given by
\begin{equation}
    r_{\text{erg}} = m + \sqrt{m^2-a^2\cos^2\theta-2m\ell-\mathcal{O}(\ell^2)} \ .
\end{equation}
This has the correct limiting behaviour as $\ell\rightarrow0$. Expanding instead around $r_{\text{erg,Kerr}}=m+\sqrt{m^2-a^2\cos^2\theta}$ yields
\begin{equation}
    r_{\text{erg}} = r_{\text{erg,Kerr}} - \frac{2m\left(r_{\text{erg,Kerr}}\right)-a^2\cos^2\theta}{\left(r_{\text{erg,Kerr}}\right)\left[r_{\text{erg,Kerr}}-\frac{a^2\cos^2\theta}{m}\right]}\,\ell + \mathcal{O}(\ell^2) \ .
\end{equation}
The surface gravity of the outer horizon in our universe is given by
\begin{equation}
    \kappa_{\text{out}} = \frac{1}{2}\,\frac{\d}{\d r}\left(\frac{\Delta_{\text{eos}}}{r^2+a^2}\right)\Bigg\vert_{r_{H}} = \frac{m\,\e^{-\ell/r_{H}}(r_{H}^3-\ell r_{H}^2-a^2r_{H}-a^2\ell)}{r_{H}(r_{H}^2+a^2)^2} \ .
\end{equation}
Imposing the extremality constraint $\kappa_{\text{out}}=0$ amounts to forcing inner and outer horizons to merge, and one recovers the condition $a^2\to a^2(r_H)$ discussed above.
Alternatively one could impose this constraint directly and find the extremal horizon location $r_H$ by solving a cubic, however for our purposes this is not informative as it gives the same qualitative information that has already been obtained.

\subsection{Killing tensor and Killing tower}\label{Killingtower}

Let us first display the relevant results in full generality for the class of ``$1$-function off-shell'' Kerr geometries. When compared with the BL coordinate system of equation~(\ref{E:eos}), the generalised line element for ``$1$-function off-shell'' Kerr simply makes the replacement $\Delta_{\text{eos}}\rightarrow \Delta = r^2+a^2-2r\,m(r)$. The contravariant metric tensor can then be written in the following form
\begin{equation}\label{contramet}
    g^{\mu\nu} = -\frac{1}{\Sigma}\begin{bmatrix} \Sigma+\frac{2r(r^2+a^2)\,m(r)}{\Delta} & 0 & 0 & \frac{2ar\,m(r)}{\Delta}\\ 0 & -\Delta & 0 & 0 \\ 0 & 0 & -1 & 0 \\ \frac{2ar\,m(r)}{\Delta} & 0 & 0 & \frac{a^2}{\Delta}-\frac{1}{\sin^2\theta}\\\end{bmatrix}^{\mu\nu} \ .
\end{equation}
With the goal of finding a nontrivial  Killing 2-tensor $K_{\mu\nu}$, satisfying $K_{(\mu\nu;\alpha)}=0$, we wish to apply the Papadopoulos--Kokkotas algorithm~\cite{Papadopoulos:2018, Papadopoulos:2020} for obtaining nontrivial Killing tensors on axisymmetric spacetimes. This algorithm is an extension of older results by Benenti and Francaviglia~\cite{Benenti:1979}, and the first step is to decompose the contravariant metric in BL-coordinates into the following general form
\begin{equation}
    g^{\mu\nu} = \frac{1}{A_1(r)+B_1(\theta)}\begin{bmatrix} A_5(r)+B_5(\theta) & 0 & 0 & A_4(r)+B_4(\theta)\\ 0 & A_2(r) &
     0 & 0\\ 0 & 0 & B_2(\theta) & 0\\ A_4(r) + B_4(\theta) & 0 & 0 & A_3(r)+B_3(\theta)\\\end{bmatrix}^{\mu\nu} \ .
\end{equation}
Equation~(\ref{contramet}) is readily interpreted to be of this form, with the explicit assignments
\begin{eqnarray}
    &A_1(r) = -r^2 \ , \quad &A_2(r) = -\Delta \ , \quad A_3(r) = \frac{a^2}{\Delta} \ , \nonumber \\
    && \nonumber \\
    &  A_4(r) = \frac{2ar\,m(r)}{\Delta} \ , \quad &A_5(r) = r^2+\frac{2r(r^2+a^2)\,m(r)}{\Delta} \ ; \nonumber \\
    && \nonumber \\
    & B_1(\theta) = -a^2\cos^2\theta \ , \quad &B_2(\theta) = -1 \ , \quad B_3(\theta) = -\frac{1}{\sin^2\theta} \ , \nonumber \\
    && \nonumber \\
    &  B_4(\theta) = 0 \ , \quad &B_5(\theta) = a^2\cos^2\theta \ .
\end{eqnarray}
Given this decomposition, the Papadopoulos--Kokkotas algorithm~\cite{Papadopoulos:2018, Papadopoulos:2020} asserts that the following yields a nontrivial contravariant Killing tensor:
\begin{equation}
    K^{\mu\nu} = \frac{1}{A_1+B_1}\begin{bmatrix} B_1A_5-A_1B_5 & 0 & 0 & B_1A_4-A_1B_4\\ 0 & A_2B_1 & 0 & 0\\ 0 & 0 & -A_1B_2 & 0\\ B_1A_4-A_1B_4 & 0 & 0 & B_1A_3-A_1B_3\\\end{bmatrix}^{\mu\nu} \ .
\end{equation}
As such, one finds the following nontrivial rank two contravariant Killing tensor for ``$1$-function off-shell'' Kerr spacetimes:
\begin{equation}
    K^{\mu\nu} = \frac{a^2\cos^2\theta}{\Sigma}\begin{bmatrix} \frac{2r(r^2+a^2)\,m(r)}{\Delta} & 0 & 0 & \frac{2ar\,m(r)}{\Delta}\\ 0 & -\Delta & 0 & 0\\ 0 & 0 & \frac{r^2}{a^2\cos^2\theta} & 0\\ \frac{2ar\,m(r)}{\Delta} & 0 & 0 & \frac{a^2}{\Delta}+\left(\frac{r}{a\sin\theta\cos\theta}\right)^2\\\end{bmatrix}^{\mu\nu} \ .
\end{equation}
Lowering the indices, one finds covariant Killing 2-tensor satisfying $K_{(\mu\nu;\alpha)}=0$ (easily verified using {\sf Maple}) in the BL coordinate basis.

Converting then to the orthonormal tetrad basis \emph{via} $K_{\hat{\mu}\hat{\nu}}=e_{\hat{\mu}}{}^{\mu}\,e_{\hat{\nu}}{}^{\nu}\,K_{\mu\nu}$, and raising the indices, gives the following
\begin{equation}
    K^{\hat{\mu}\hat{\nu}} = \begin{bmatrix}-a^2\cos^2\theta & 0 & 0 & 0\\ 0 & a^2\cos^2\theta & 0 & 0\\ 0 & 0 & -r^2 & 0\\ 0 & 0 & 0 & -r^2\\\end{bmatrix}^{\hat{\mu}\hat{\nu}} \ .
\end{equation}
Notice that in the tetrad basis this is identical to the nontrivial Killing 2-tensor for Kerr spacetime. Specifically, notice that it is \emph{independent} of the  mass function $m(r)$. As such, the entire family of ``$1$-function off-shell'' Kerr geometries inherits the same Killing tensor as Kerr spacetime. Notably, both the Ricci tensor and Killing tensor are diagonal in this tetrad basis, and as such the commutator $[R,K]^{\hat{\mu}}{}_{\hat{\nu}}$ will vanish; it has been recently proven that this constraint is sufficient to conclude that the Klein--Gordon equation is separable on the background spacetime~\cite{PGLT2}. As such, the eye of the storm is amenable to a standard spin zero quasi-normal modes analysis (invoke the inverse Cowling effect, assume a separable wave form, and use your favourite numerical technique to approximate the ringdown signal). The same can be said for all candidate geometries in the class of ``$1$-function off-shell'' Kerr~\cite{Frolov:2017, Carter:1968a, Carter:1968b}.

Furthermore, it is straightforward (\emph{e.g.}, using {\sf Maple}) to verify that the following two-form square-root of the Killing tensor is a genuine Killing--Yano tensor, satisfying the Killing--Yano equation $f_{\hat{\mu}(\hat{\nu};\hat{\alpha})}=0$:
\begin{equation}
    f^{\hat{\mu}\hat{\nu}} = \begin{bmatrix} 0 & a\cos\theta & 0 & 0\\ -a\cos\theta & 0 & 0 & 0\\ 0 & 0 & 0 & -r\\ 0 & 0 & r & 0\\\end{bmatrix}^{\hat{\mu}\hat{\nu}} \ .
\end{equation}
It is straightforward to check that $K^{\hat{\mu}\hat{\nu}} = -f^{\hat{\mu}\hat{\alpha}}\;\eta_{\hat{\alpha}\hat{\beta}}\;f^{\hat{\beta}\hat{\nu}}$. The ``principal tensor''~\cite{Frolov:2017} is then simply the Hodge dual of this two-form, and in full generality the family of ``$1$-function off-shell'' Kerr geometries possesses the full ``Killing tower''~\cite{Frolov:2017} of Killing tensor, Killing--Yano tensor, and principal tensor. Separability of the Hamilton--Jacobi equations is guaranteed by the existence of $K_{\mu\nu}$ and the associated Carter constant, and in reference~\citenum{Ghosh:2020a} the geodesics for the photon ring are computed. Notably, the eye of the storm is able to be delineated from Kerr, and results from reference~\citenum{Ghosh:2020b} conclude that the data from the image of M87 provided by the EHT does not exclude the eye of the storm from being astrophysically viable. This is yet another highly desirable feature of the eye of the storm geometry.

The spacetime is also amenable to straightforward calculation of the black hole shadow~\cite{Ghosh:2016, Tsukamoto:2017}. These calculations further demonstrate that the geometry falls within experimental constraints provided by the EHT. Furthermore, the fact that the eye of the storm falls within this class of ``$1$-function off-shell" Kerr geometries implies that Maxwell's equations also separate on the background spacetime; this is confirmed by the proof given in Appendix A of reference~\cite{Franzin:2021max}. We conjecture that the equations governing the spin two polar and axial modes will also be separable on this geometry.

\section{Stress-energy and energy conditions}

Recall that the Einstein tensor in an orthonormal basis is given by
\begin{eqnarray}
    G^{\hat{\mu}}{}_{\hat{\nu}} = -\frac{2\ell m\,\e^{-\ell/r}}{\Sigma^2}\,\text{diag}\left(1,1,\Xi-1,\Xi-1\right) \ ,
\end{eqnarray}
where we have used $\Xi=\frac{\ell\Sigma}{2r^3}$. We wish to fix the geometrodynamics by interpreting the spacetime through the lens of standard GR. As such, coupling the geometry to the Einstein equations, we have
\begin{equation}
    \frac{1}{8\pi}G^{\hat{\mu}}{}_{\hat{\nu}} = T^{\hat{\mu}}{}_{\hat{\nu}} = \text{diag}(-\rho,p_{r},p_{t},p_{t}) \ .
\end{equation}
Due to the fact that $-\rho=p_{r}$ this equation holds globally in the geometry, regardless of whether one is outside (inside) the outer (inner) horizons, or trapped in between them. The fact that the Einstein tensor is diagonal in an orthonormal basis implies that the stress-energy tensor is Hawking--Ellis type I~\cite{Hawking:1973, Martin-Moruno:2017, Martin-Moruno:2018, Martin-Moruno:2021}. This leads to the following specific stress-energy components:
\begin{eqnarray}
    \rho &=& -p_{r} = \frac{\ell m\,\e^{-\ell/r}}{4\pi\Sigma^2} \ , \nonumber \\
    && \nonumber \\
    p_{t} &=& \frac{\ell m\,\e^{-\ell/r}}{4\pi\Sigma^2}\left(1-\Xi\right) \ .
\end{eqnarray}
An extremely desirable feature of the ``eye of the storm'' spacetime is its relationship with the classical energy conditions associated with GR. In view of $\ell>0$, one trivially globally satisfies $\rho>0$. The radial null energy condition (NEC) is trivially satisfied since $\rho+p_{r}=0$ globally. Analysing the transverse NEC:
\begin{equation}
    \rho + p_{t} = \frac{\ell m\,\e^{-\ell/r}}{4\pi\Sigma^2}\left(2-\Xi\right) \ .
\end{equation}
This changes sign when $\Xi=2$, or when $\frac{\Sigma}{r^3}=\frac{4}{\ell}$. On the equatorial plane this is when $r=\frac{\ell}{4}$. If $\frac{\Sigma}{r^3}>\frac{4}{\ell}$, the transverse NEC is violated, whilst if $\frac{\Sigma}{r^3}<\frac{4}{\ell}$, it is satisfied. For the equatorial plane, the violated region is when $r<\frac{\ell}{4}$.

Let us look at the strong energy condition (SEC), which implies $\rho+p_{r}+2p_{t}>0$:
\begin{equation}
    \rho + p_{r} + 2p_{t} = 2p_{t} = \frac{\ell m\,\e^{-\ell/r}}{2\pi\Sigma^2}\left(1-\Xi\right) \ .
\end{equation}
This changes sign when $\Xi=1$, or when $\frac{\Sigma}{r^3}=\frac{2}{\ell}$. If $\frac{\Sigma}{r^3}>\frac{2}{\ell}$, the SEC is violated, whilst if $\frac{\Sigma}{r^3}<\frac{2}{\ell}$, the SEC is satisfied. For the equatorial plane the violated region is whenever $r<\frac{\ell}{2}$.

Given the freedom to choose the suppression parameter $\ell$, this means that all of the energy-condition-violating physics can be forced into an arbitrarily small region in the deep core. One can conceive of three sensible categories of relativist in the present day:\enlargethispage{20pt}
\begin{itemize}
    \item Those who believe that GR holds \emph{everywhere}, other than at a distance scale where a mature and phenomenologically verifiable theory of quantum gravity must necessarily take over.
    \item Those who believe that GR can only be believed in regions external to any Cauchy horizon(s).
    \item Those who believe that GR only holds outside \emph{any} horizon full stop.
\end{itemize}
%

Regardless of one's personal subscription, the freedom to scale $\ell$ as required means all of the energy-condition-violating physics can be readily pushed into a region where GR is no longer an appropriate theory. Notably, no exotic matter is required in the exterior region of the spacetime. In the domain of outer communication, we have manifest satisfaction of all of the classical energy conditions. This is consistent with astrophysical observations, and is an extremely desirable feature of eye of the storm spacetime when compared with the remaining literature concerning rotating RBHs; for instance in the spacetimes explored in references~\citenum{Mazza:2021, Franzin:2021} one has global violation of the NEC.

\section{Conclusions}\label{S:discussion}

We have defined the class of ``$1$-function off-shell'' Kerr geometries, and demonstrated the general existence of the full Killing tower for the geometries within it. Within this class, we have selected the most desirable candidate spacetime, the eye of the storm, according to a set of carefully chosen theoretically and experimentally motivated metric construction criteria. This spacetime models a rotating regular black hole with asymptotically Minkowski core, is asymptotically Kerr for large $r$, manifestly satisfies all of the standard energy conditions of GR in both the region of theoretical validity and the region of experimental validity, has integrable geodesics in principle, and has the property of separability of the Klein--Gordon equation. The eye of the storm is also the most mathematically tractable rotating RBH in the current literature, and is readily amenable to the extraction of astrophysical observables falsifiable/verifiable by the experimental community.

Separability of both the Klein--Gordon equation and Maxwell's equations leads us to conjecture that the equations governing the spin two polar and axial modes will also separate on this spacetime. Verifying this is an important topic for future research. Extracting the full family of geodesics for test particles in the spacetime is also an important calculation. Ultimately, one should calculate as far as is possible the geodesics in full generality, and probe the geometry for quasi-normal modes analysis.

\section*{Acknowledgements}

\enlargethispage{40pt}
AS was supported by a Victoria University of Wellington PhD scholarship, and was also indirectly supported by the Marsden Fund, \emph{via} a grant administered by the Royal Society of New Zealand.\\
AS would also like to thank Ratu Mataira for useful conversations and discussion.\\
MV was directly supported by the Marsden Fund, \emph{via} a grant administered by the
Royal Society of New Zealand.



\begin{thebibliography}{99}
\newcommand{\arXiv}[1]{arXiv:~{\href{https://arxiv.org/abs/#1}{\color{blue}#1}}}

\bibitem{LIGO1}
See \url{https://www.ligo.caltech.edu/page/detection-companion-papers} for a collection of detection papers from LIGO. \\
See also \url{https://pnp.ligo.org/ppcomm/Papers.html} for a complete list of publications from the LIGO Scientific Collaboration and Virgo Collaboration.


\bibitem{LIGO2}
See, for example, \href{https://en.wikipedia.org/wiki/List_of_gravitational_wave_observations}{wikipedia.org/List\_of\_gravitational\_wave\_observations} for a list of current (December 2021) gravitational wave observations.


\bibitem{EHT1}
K.~Akiyama \textit{et al.} [Event Horizon Telescope],
``First M87 Event Horizon Telescope Results. I. The Shadow of the Supermassive Black Hole'',
Astrophys. J. Lett. \textbf{875} (2019), L1,
\doi{10.3847/2041-8213/ab0ec7},
[\href{https://arxiv.org/ftp/arxiv/papers/1906/1906.11238.pdf}{arXiv:1906.11238} [astro-ph.GA]].

\bibitem{EHT2}
K.~Akiyama \textit{et al.} [Event Horizon Telescope],
``First M87 Event Horizon Telescope Results. II. Array and Instrumentation'',
Astrophys. J. Lett. \textbf{875} (2019) no.1, L2,
\doi{10.3847/2041-8213/ab0c96},
[\href{https://arxiv.org/ftp/arxiv/papers/1906/1906.11239.pdf}{arXiv:1906.11239} [astro-ph.IM]].

\bibitem{EHT3}
K.~Akiyama \textit{et al.} [Event Horizon Telescope],
``First M87 Event Horizon Telescope Results. III. Data Processing and Calibration'',
Astrophys. J. Lett. \textbf{875} (2019) no.1, L3,
\doi{10.3847/2041-8213/ab0c57},
[\href{https://arxiv.org/ftp/arxiv/papers/1906/1906.11240.pdf}{arXiv:1906.11240} [astro-ph.GA]].

\bibitem{EHT4}
K.~Akiyama \textit{et al.} [Event Horizon Telescope],
``First M87 Event Horizon Telescope Results. IV. Imaging the Central Supermassive Black Hole'',
Astrophys. J. Lett. \textbf{875} (2019) no.1, L4,
\doi{10.3847/2041-8213/ab0e85},
[\href{https://arxiv.org/ftp/arxiv/papers/1906/1906.11241.pdf}{arXiv:1906.11241} [astro-ph.GA]].

\bibitem{EHT5}
K.~Akiyama \textit{et al.} [Event Horizon Telescope],
``First M87 Event Horizon Telescope Results. V. Physical Origin of the Asymmetric Ring'',
Astrophys. J. Lett. \textbf{875} (2019) no.1, L5,
\doi{10.3847/2041-8213/ab0f43},
[\href{https://arxiv.org/ftp/arxiv/papers/1906/1906.11242.pdf}{arXiv:1906.11242} [astro-ph.GA]].

\bibitem{EHT6}
K.~Akiyama \textit{et al.} [Event Horizon Telescope],
``First M87 Event Horizon Telescope Results. VI. The Shadow and Mass of the Central Black Hole'',
Astrophys. J. Lett. \textbf{875} (2019) no.1, L6,
\doi{10.3847/2041-8213/ab1141},\\{}
[\href{https://arxiv.org/ftp/arxiv/papers/1906/1906.11243.pdf}{arXiv:1906.11243} [astro-ph.GA]].


\bibitem{LISA}
E.~Barausse, E.~Berti, T.~Hertog, S.~A.~Hughes, P.~Jetzer, P.~Pani, T.~P.~Sotiriou, N.~Tamanini, H.~Witek, K.~Yagi, N.~Yunes, \emph{et al.},\\
``Prospects for Fundamental Physics with LISA'',\\
Gen. Rel. Grav. \textbf{52} (2020) no.8, 81.
\doi{10.1007/s10714-020-02691-1},
[\href{https://arxiv.org/abs/2001.09793}{arXiv:2001.09793} [gr-qc]].


\bibitem{science-book}
V.~Kalogera, B.~S.~Sathyaprakash, M.~Bailes, M.~A.~Bizouard, A.~Buonanno, A.~Burrows, M.~Colpi, M.~Evans, S.~Fairhurst and S.~Hild, \textit{et al.}
``The Next Generation Global Gravitational Wave Observatory: The Science Book'',
[\arXiv{2111.06990} [gr-qc]].


\bibitem{Eiroa:2010}
E.~F.~Eiroa and C.~M.~Sendra,
``Gravitational lensing by a regular black hole'',
Class. Quant. Grav. \textbf{28} (2011), 085008,
\doi{10.1088/0264-9381/28/8/085008},
[\href{https://arxiv.org/pdf/1011.2455.pdf}{arXiv:1011.2455} [gr-qc]].

\bibitem{Flachi:2012}
A.~Flachi and J.~P.~S.~Lemos,
``Quasinormal modes of regular black holes'',\\
Phys. Rev. D \textbf{87} (2013) no.2, 024034,
\doi{10.1103/PhysRevD.87.024034},
[\href{https://arxiv.org/pdf/1211.6212.pdf}{arXiv:1211.6212} [gr-qc]].

\bibitem{Abdujabbarov:2016}
A.~Abdujabbarov, M.~Amir, B.~Ahmedov and S.~G.~Ghosh,\\
``Shadow of rotating regular black holes'',
Phys. Rev. D \textbf{93} (2016) no.10, 104004,
\doi{10.1103/PhysRevD.93.104004},
[\href{https://arxiv.org/pdf/1604.03809.pdf}{arXiv:1604.03809} [gr-qc]].

\bibitem{Carballo-Rubio:2018}
R.~Carballo-Rubio, F.~Di Filippo, S.~Liberati, C.~Pacilio and M.~Visser,\\
``On the viability of regular black holes'',
JHEP \textbf{07} (2018), 023,
\doi{10.1007/JHEP07(2018)023},
[\href{https://arxiv.org/pdf/1805.02675.pdf}{arXiv:1805.02675} [gr-qc]].

\bibitem{Carballo-Rubio:2019a}
R.~Carballo-Rubio, F.~Di Filippo, S.~Liberati and M.~Visser,\\
``Opening the Pandora\textquoteright{}s box at the core of black holes'',\\
Class. Quant. Grav. \textbf{37} (2020) no.14, 14,
\doi{10.1088/1361-6382/ab8141},
[\href{https://arxiv.org/pdf/1908.03261.pdf}{arXiv:1908.03261} [gr-qc]].

\bibitem{Carballo-Rubio:2019b}
R.~Carballo-Rubio, F.~Di Filippo, S.~Liberati and M.~Visser,\\
``Geodesically complete black holes'',
Phys. Rev. D \textbf{101} (2020), 084047,
\doi{10.1103/PhysRevD.101.084047},
[\href{https://arxiv.org/pdf/1911.11200.pdf}{arXiv:1911.11200} [gr-qc]].



\bibitem{Dai:2019}
D.~C.~Dai and D.~Stojkovic,
``Observing a Wormhole'',\\
Phys. Rev. D \textbf{100} (2019) no.8, 083513,
\doi{10.1103/PhysRevD.100.083513},
[\href{https://arxiv.org/pdf/1910.00429.pdf}{arXiv:1910.00429} [gr-qc]].

\bibitem{Cramer:1994}
J.~G.~Cramer, R.~L.~Forward, M.~S.~Morris, M.~Visser, G.~Benford and G.~A.~Landis,
``Natural wormholes as gravitational lenses'',
Phys. Rev. D \textbf{51} (1995), 3117-3120
\doi{10.1103/PhysRevD.51.3117},
[\arXiv{astro-ph/9409051} [astro-ph]].

\bibitem{Simonetti:2020}
J.~H.~Simonetti, M.~J.~Kavic, D.~Minic, D.~Stojkovic and D.~C.~Dai,\\
``Sensitive searches for wormholes'',
Phys. Rev. D \textbf{104} (2021) no.8, L081502,
\doi{10.1103/PhysRevD.104.L081502},
[\href{https://arxiv.org/pdf/2007.12184.pdf}{arXiv:2007.12184} [gr-qc]].

\bibitem{Berry:2020}
T.~Berry, A.~Simpson and M.~Visser,
``Photon spheres, ISCOs, and OSCOs: Astrophysical observables for regular black holes with asymptotically Minkowski cores'',
Universe \textbf{7} (2020) no.1, 2,
\doi{10.3390/universe7010002},\\{}
[\href{https://arxiv.org/pdf/2008.13308.pdf}{arXiv:2008.13308} [gr-qc]].

\bibitem{Carballo-Rubio:2021a}
R.~Carballo-Rubio, F.~Di Filippo and S.~Liberati,\\
``Hearts of Darkness: the inside out probing of black holes'', 
IJMPD (in press),
\doi{10.1142/S0218271821420244},
[\href{https://arxiv.org/pdf/2106.01530.pdf}{arXiv:2106.01530} [gr-qc]].

\bibitem{Carballo-Rubio:2021b}
R.~Carballo-Rubio, F.~Di Filippo, S.~Liberati and M.~Visser,
``Geodesically complete black holes in Lorentz-violating gravity'',
[\arXiv{2111.03113} [gr-qc]].

\bibitem{Bronnikov:2021}
K.~A.~Bronnikov, R.~A.~Konoplya and T.~D.~Pappas,
``General parametrization of wormhole spacetimes and its application to shadows and quasinormal modes'',
Phys. Rev. D \textbf{103} (2021) no.12, 124062,
\doi{10.1103/PhysRevD.103.124062},
[\href{https://arxiv.org/pdf/2102.10679.pdf}{arXiv:2102.10679} [gr-qc]].

\bibitem{Churilova:2021}
M.~S.~Churilova, R.~A.~Konoplya, Z.~Stuchlik and A.~Zhidenko,
``Wormholes without exotic matter: quasinormal modes, echoes and shadows'',
JCAP \textbf{10} (2021), 010,
\doi{10.1088/1475-7516/2021/10/010},
[\href{https://arxiv.org/pdf/2107.05977.pdf}{arXiv:2107.05977} [gr-qc]].

\bibitem{Bambi:2021}
C.~Bambi and D.~Stojkovic,
``Astrophysical Wormholes'',
Universe \textbf{7} (2021) no.5, 136,
\doi{10.3390/universe7050136},
[\href{https://arxiv.org/pdf/2105.00881.pdf}{arXiv:2105.00881} [gr-qc]].

\bibitem{Simpson:2021biv}
A.~Simpson,
``Ringing of the Regular Black Hole with Asymptotically Minkowski Core'',
Universe \textbf{7} (2021) no.11, 418,
\doi{10.3390/universe7110418},\\{}
[\href{https://arxiv.org/pdf/2109.11878.pdf}{arXiv:2109.11878} [gr-qc]].

\bibitem{PGLT1}
J.~Baines, T.~Berry, A.~Simpson and M.~Visser,
``Painlev\'e\textendash{}Gullstrand form of the Lense\textendash{}Thirring Spacetime'',
Universe \textbf{7} (2021) no.4, 105,
\doi{10.3390/universe7040105},
[\arXiv{2006.14258} [gr-qc]].

\bibitem{Visser:epoch}
M.~Visser,
``Energy conditions in the epoch of galaxy formation'',
Science \textbf{276} (1997), 88--90,
\doi{10.1126/science.276.5309.88},
[\arXiv{1501.01619} [gr-qc]].

\bibitem{Visser:epoch-prd}
M.~Visser,
``General relativistic energy conditions: The Hubble expansion in the epoch of galaxy formation'',
Phys. Rev. D \textbf{56} (1997), 7578--7587,
\doi{10.1103/PhysRevD.56.7578},
[\arXiv{gr-qc/9705070} [gr-qc]].

\bibitem{Visser:cosmo99}
M.~Visser and C.~Barcel\'o,
``Energy conditions and their cosmological implications'',
\doi{10.1142/9789812792129\_0014},
[\arXiv{gr-qc/0001099} [gr-qc]].


\bibitem{Bardeen:1968}
    J.~M.~Bardeen, ``Non-singular general relativistic gravitational collapse'',
    Abstracts of the 5th international conference on gravitation and the theory of relativity (GR5), 
    eds. V.~A.~Fock \emph{et al.} (Tbilisi University Press, Tblisi, Georgia,  former USSR, 1968), 
    pages 174--175.
    
\bibitem{Ayon-Beato:2000}
E.~Ayon-Beato and A.~Garcia,
``The Bardeen model as a nonlinear magnetic monopole'',
Phys. Lett. B \textbf{493} (2000), 149--152,
\doi{10.1016/S0370-2693(00)01125-4},
[\href{https://arxiv.org/pdf/gr-qc/0009077.pdf}{arXiv:gr-qc/0009077} [gr-qc]].

\bibitem{Hayward:2005}
S.~A.~Hayward,
``Formation and evaporation of regular black holes'',\\
Phys. Rev. Lett. \textbf{96} (2006), 031103,
\doi{10.1103/PhysRevLett.96.031103},
[\href{https://arxiv.org/pdf/gr-qc/0506126.pdf}{arXiv:gr-qc/0506126} [gr-qc]].

\bibitem{Bronnikov:2005}
K.~A.~Bronnikov and J.~C.~Fabris,
``Regular phantom black holes'',\\
Phys. Rev. Lett. \textbf{96} (2006), 251101,
\doi{10.1103/PhysRevLett.96.251101},
[\href{https://arxiv.org/abs/gr-qc/0511109}{arXiv:gr-qc/0511109} [gr-qc]].
    
\bibitem{Bambi:2013}
C.~Bambi and L.~Modesto,
``Rotating regular black holes'',
Phys. Lett. B \textbf{721} (2013), 329--334,
\doi{10.1016/j.physletb.2013.03.025},
[\href{https://arxiv.org/pdf/1302.6075.pdf}{arXiv:1302.6075} [gr-qc]].

\bibitem{Li:2013}
Z.~Li and C.~Bambi,
``Measuring the Kerr spin parameter of regular black holes from their shadow'',
JCAP \textbf{01} (2014), 041,
\doi{10.1088/1475-7516/2014/01/041},
[\arXiv{1309.1606} [gr-qc]].

\bibitem{Jusufi:2017}
K.~Jusufi and A.~\"Ovg\"un,
``Gravitational Lensing by Rotating Wormholes'',\\
Phys. Rev. D \textbf{97} (2018) no.2, 024042,
\doi{10.1103/PhysRevD.97.024042},
[\arXiv{1708.06725} [gr-qc]].

\bibitem{Jusufi:2019}
K.~Jusufi, M.~Jamil, H.~Chakrabarty, Q.~Wu, C.~Bambi and A.~Wang,
``Rotating regular black holes in conformal massive gravity'',
Phys. Rev. D \textbf{101} (2020) no.4, 044035,
\doi{10.1103/PhysRevD.101.044035},
[\arXiv{1911.07520} [gr-qc]].

\bibitem{Jusufi:2020}
K.~Jusufi, M.~Azreg-A\"\i{}nou, M.~Jamil, S.~W.~Wei, Q.~Wu and A.~Wang,
``Quasinormal modes, quasiperiodic oscillations, and the shadow of rotating regular black holes in nonminimally coupled Einstein-Yang-Mills theory'',
Phys. Rev. D \textbf{103} (2021) no.2, 024013,
\doi{10.1103/PhysRevD.103.024013},
[\arXiv{2008.08450} [gr-qc]].


\bibitem{Herdeiro:2016}
C.~Herdeiro, E.~Radu and H.~R\'unarsson,
``Kerr black holes with Proca hair'',\\
Class. Quant. Grav. \textbf{33} (2016) no.15, 154001,
\doi{10.1088/0264-9381/33/15/154001},
[\arXiv{1603.02687} [gr-qc]].

\bibitem{Frolov:2014}
V.~P.~Frolov,
``Information loss problem and a `black hole' model with a closed apparent horizon'',
JHEP \textbf{05} (2014), 049,
\doi{10.1007/JHEP05(2014)049},
[\href{https://arxiv.org/pdf/1402.5446.pdf}{arXiv:1402.5446} [hep-th]].

\bibitem{Ghosh:2014}
S.~G.~Ghosh, ``A nonsingular rotating black hole'',
Eur. Phys. J. C \textbf{75} (2015) no.11, 532,
\doi{10.1140/epjc/s10052-015-3740-y},
[\href{https://arxiv.org/pdf/1408.5668.pdf}{arXiv:1408.5668} [gr-qc]].
    
\bibitem{Neves:2014}
J.~C.~S.~Neves and A.~Saa,
``Regular rotating black holes and the weak energy condition'',
Phys. Lett. B \textbf{734} (2014), 44--48,
\doi{10.1016/j.physletb.2014.05.026},
[\href{https://arxiv.org/pdf/1402.2694.pdf}{arXiv:1402.2694} [gr-qc]].

\bibitem{Toshmatov:2014}
B.~Toshmatov, B.~Ahmedov, A.~Abdujabbarov and Z.~Stuchlik,
``Rotating Regular Black Hole Solution'',
Phys. Rev. D \textbf{89} (2014) no.10, 104017,
\doi{10.1103/PhysRevD.89.104017},
[\href{https://arxiv.org/pdf/1404.6443.pdf}{arXiv:1404.6443} [gr-qc]].

\bibitem{quadratic}
V.~K.~Tinchev, ``The Shadow of Generalized Kerr Black Holes with Exotic Matter'',
Chin. J. Phys. \textbf{53} (2015), 110113,
\doi{10.6122/CJP.20150810},\\{}
[\arXiv{1512.09164} [gr-qc]].
    
\bibitem{Fan:2016}
Z.~Y.~Fan and X.~Wang,
``Construction of Regular Black Holes in General Relativity'',
Phys. Rev. D \textbf{94} (2016) no.12, 124027,
\doi{10.1103/PhysRevD.94.124027},
[\href{https://arxiv.org/pdf/1610.02636.pdf}{arXiv:1610.02636} [gr-qc]].

\bibitem{Toshmatov:2017a}
B.~Toshmatov, Z.~Stuchl\'\i{}k and B.~Ahmedov,
``Generic rotating regular black holes in general relativity coupled to nonlinear electrodynamics'',
Phys. Rev. D \textbf{95} (2017) no.8, 084037,
\doi{10.1103/PhysRevD.95.084037},
[\href{https://arxiv.org/pdf/1704.07300.pdf}{arXiv:1704.07300} [gr-qc]].

\bibitem{Toshmatov:2017b}
B.~Toshmatov, Z.~Stuchl\'\i{}k and B.~Ahmedov,
``Note on the character of the generic rotating charged regular black holes in general relativity coupled to nonlinear electrodynamics'',
[\href{https://arxiv.org/pdf/1712.04763.pdf}{arXiv:1712.04763} [gr-qc]].

\bibitem{Simpson:2018}
A.~Simpson and M.~Visser,
``Black-bounce to traversable wormhole'',
JCAP \textbf{02} (2019), 042,
\doi{10.1088/1475-7516/2019/02/042},
[\href{https://arxiv.org/pdf/1812.07114.pdf}{arXiv:1812.07114} [gr-qc]].

\bibitem{Simpson:2019}
A.~Simpson, P.~Mart\'{i}n-Moruno and M.~Visser,
``Vaidya spacetimes, black-bounces, and traversable wormholes'',
Class. Quant. Grav. \textbf{36} (2019) no.14, 145007,
\doi{10.1088/1361-6382/ab28a5},
[\href{https://arxiv.org/pdf/1902.04232.pdf}{arXiv:1902.04232} [gr-qc]].

\bibitem{Lobo:2020a}
F.~S.~N.~Lobo, M.~E.~Rodrigues, M.~V.~d.~S.~Silva, A.~Simpson and M.~Visser,\\
``Novel black-bounce spacetimes: wormholes, regularity, energy conditions, and causal structure'',
Phys. Rev. D \textbf{103} (2021) no.8, 084052,
\doi{10.1103/PhysRevD.103.084052},
[\href{https://arxiv.org/pdf/2009.12057.pdf}{arXiv:2009.12057} [gr-qc]].

\bibitem{Simpson:2020}
A. Simpson and M. Visser,
``Regular black holes with asymptotically Minkowski cores'',
Universe \textbf{6} (2020) 8,
\doi{10.3390/universe6010008},
[\href{https://arxiv.org/abs/1911.01020}{arXiv:1911.01020} [gr-qc]].

\bibitem{Brahma:2020}
S.~Brahma, C.~Y.~Chen and D.~h.~Yeom, ``Testing Loop Quantum Gravity from Observational Consequences of Nonsingular Rotating Black Holes'',
Phys. Rev. Lett. \textbf{126} (2021) no.18, 181301,
\doi{10.1103/PhysRevLett.126.181301},\\{}
[\href{https://arxiv.org/pdf/2012.08785.pdf}{arXiv:2012.08785} [gr-qc]].

\bibitem{Mazza:2021}
J.~Mazza, E.~Franzin and S.~Liberati,
``A novel family of rotating black hole mimickers'',
JCAP \textbf{04} (2021), 082,
\doi{10.1088/1475-7516/2021/04/082},
[\href{https://arxiv.org/pdf/2102.01105.pdf}{arXiv:2102.01105} [gr-qc]].

\bibitem{Franzin:2021}
E.~Franzin, S.~Liberati, J.~Mazza, A.~Simpson and M.~Visser,
``Charged black-bounce spacetimes'',
JCAP \textbf{07} (2021), 036,
\doi{10.1088/1475-7516/2021/07/036},
[\href{https://arxiv.org/pdf/2104.11376.pdf}{arXiv:2104.11376} [gr-qc]].


\bibitem{Frolov:2017}
V.~Frolov, P.~Krtous and D.~Kubiznak,
``Black holes, hidden symmetries, and complete integrability'',
Living Rev. Rel. \textbf{20} (2017) no.1, 6,
\doi{10.1007/s41114-017-0009-9},
[\href{https://arxiv.org/pdf/1705.05482.pdf}{arXiv:1705.05482} [gr-qc]].

\bibitem{Carter:1968a}
B.~Carter,
``Global structure of the Kerr family of gravitational fields'',
Phys. Rev. \textbf{174} (1968), 1559--1571,
\doi{10.1103/PhysRev.174.1559}.

\bibitem{Carter:1968b}
B.~Carter,
``Hamilton-Jacobi and Schr\"odinger separable solutions of Einstein's equations'',
Commun. Math. Phys. \textbf{10} (1968) no.4, 280--310,
\doi{10.1007/BF03399503}.

\bibitem{Kroon}
J.~A.~Valiente Kroon,
``On the nonexistence of conformally flat slices in the Kerr and other stationary space-times'',
Phys. Rev. Lett. \textbf{92} (2004), 041101,
\doi{10.1103/PhysRevLett.92.041101},
[\arXiv{gr-qc/0310048} [gr-qc]].

\bibitem{Darboux}
J.~Baines, T.~Berry, A.~Simpson and M.~Visser,
``Darboux diagonalization of the spatial 3-metric in Kerr spacetime'',
Gen. Rel. Grav. \textbf{53} (2021) no.1, 3,
\doi{10.1007/s10714-020-02765-0},
[\arXiv{2009.01397} [gr-qc]].

\bibitem{MTW}
C.~W.~Misner, K.~S.~Thorne and J.~A.~Wheeler,
\emph{Gravitation}, (Freeman, San~Francisco, 1973).

\bibitem{Wald}
R.~M.~Wald,
\emph{General Relativity}, (U Chicago Press, Chicago, 1984),
\doi{10.7208/chicago/9780226870373.001.000}.

\bibitem{Lorentzianwormholes}
M.~Visser,
\emph{Lorentzian wormholes: From Einstein to Hawking},\\
(AIP Press; now Springer, Reading, 1996).

\bibitem{Jacobson:2007}
T.~Jacobson,
``When is $g_{tt}\, g_{rr} = -1$?'',
Class. Quant. Grav. \textbf{24} (2007), 5717--5719,
\doi{10.1088/0264-9381/24/22/N02},
[\arXiv{0707.3222} [gr-qc]].


\bibitem{Eichhorn:2021a}
A.~Eichhorn and A.~Held,
``Image features of spinning regular black holes based on a locality principle'',
Eur. Phys. J. C \textbf{81} (2021) no.10, 933,
\doi{10.1140/epjc/s10052-021-09716-2},
[\href{https://arxiv.org/abs/2103.07473}{arXiv:2103.07473} [gr-qc]].

\bibitem{Eichhorn:2021b}
A.~Eichhorn and A.~Held,
``From a locality-principle for new physics to image features of regular spinning black holes with disks'',
JCAP \textbf{05} (2021), 073,
\doi{10.1088/1475-7516/2021/05/073},
[\href{https://arxiv.org/pdf/2103.13163.pdf}{arXiv:2103.13163} [gr-qc]].


\bibitem{Doran}
C.~Doran,
``A New form of the Kerr solution'',
Phys. Rev. D \textbf{61} (2000), 067503,
\doi{10.1103/PhysRevD.61.067503},
[\arXiv{gr-qc/9910099} [gr-qc]].

\bibitem{brief-introduction}
M.~Visser,
``The Kerr spacetime: A Brief introduction'',
[\arXiv{0706.0622} [gr-qc]].
Published in~\cite{Kerr-book}.

\bibitem{Kerr-book}
D.~L.~Wiltshire, M.~Visser and S.~M.~Scott, editors,
\emph{The Kerr spacetime: Rotating black holes in general relativity},
(Cambridge, England, 2009).

\bibitem{unit-lapse}
J.~Baines, T.~Berry, A.~Simpson and M.~Visser,
``Unit-lapse versions of the Kerr spacetime'',
Class. Quant. Grav. \textbf{38} (2021) no.5, 055001,
\doi{10.1088/1361-6382/abd071},
[\arXiv{2008.03817} [gr-qc]].


\bibitem{Papadopoulos:2018}
G.~O.~Papadopoulos and K.~D.~Kokkotas,
``Preserving Kerr symmetries in deformed spacetimes'',
Class. Quant. Grav. \textbf{35} (2018) no.18, 185014,
\doi{10.1088/1361-6382/aad7f4},
[\href{https://arxiv.org/pdf/1807.08594.pdf}{arXiv:1807.08594} [gr-qc]].

\bibitem{Papadopoulos:2020}
G.~O.~Papadopoulos and K.~D.~Kokkotas,
``On Kerr black hole deformations admitting a Carter constant and an invariant criterion for the separability of the wave equation'',
Gen. Rel. Grav. \textbf{53} (2021) no.2, 21,
\doi{10.1007/s10714-021-02795-2},
[\href{https://arxiv.org/abs/2007.12125}{arXiv:2007.12125} [gr-qc]].

\bibitem{Benenti:1979}
S.~Benenti and M.~Francaviglia, ``Remarks on Certain Separability Structures and Their Applications to General Relativity", 
General Relativity and Gravitation \textbf{10} (1979), 79--92,
\doi{10.1007/BF00757025}.


\bibitem{PGLT2}
J.~Baines, T.~Berry, A.~Simpson and M.~Visser,
``Killing tensor and Carter constant for Painleve-Gullstrand form of Lense-Thirring spacetime’’,
Universe {\bf 7 \#12}  (2021) 473; 
\doi{10.3390/universe7120473},
[\arXiv{2110.01814} [gr-qc]].


\bibitem{Ghosh:2020a}
R.~Kumar and S.~G.~Ghosh, ``Photon ring structure of rotating regular black holes and no-horizon spacetimes'',
Class. Quant. Grav. \textbf{38} (2021) no.8, 8,
\doi{10.1088/1361-6382/abdd48},
[\href{https://arxiv.org/pdf/2004.07501.pdf}{arXiv:2004.07501} [gr-qc]].

\bibitem{Ghosh:2020b}
R.~Kumar, A.~Kumar and S.~G.~Ghosh, ``Testing Rotating Regular Metrics as Candidates for Astrophysical Black Holes'',
Astrophys. J. \textbf{896} (2020) no.1, 89, \doi{10.3847/1538-4357/ab8c4a},
[\href{https://arxiv.org/pdf/2006.09869.pdf}{arXiv:2006.09869} [gr-qc]].

\bibitem{Ghosh:2016}
M.~Amir and S.~G.~Ghosh, ``Shapes of rotating nonsingular black hole shadows'',
Phys. Rev. D \textbf{94} (2016) no.2, 024054,
\doi{10.1103/PhysRevD.94.024054},
[\href{https://arxiv.org/pdf/1603.06382.pdf}{arXiv:1603.06382} [gr-qc]].

\bibitem{Tsukamoto:2017}
N.~Tsukamoto, ``Black hole shadow in an asymptotically-flat, stationary, and axisymmetric spacetime: The Kerr-Newman and rotating regular black holes'',
Phys. Rev. D \textbf{97} (2018) no.6, 064021, \doi{10.1103/PhysRevD.97.064021},\\{}
[\href{https://arxiv.org/pdf/1708.07427.pdf}{arXiv:1708.07427} [gr-qc]].

\bibitem{Franzin:2021max}
E.~Franzin, S.~Liberati and M.~Oi, ``Superradiance in Kerr-like black holes'',
Phys. Rev. D \textbf{103} (2021) no.10, 104034, \doi{10.1103/PhysRevD.103.104034},
[\href{https://arxiv.org/abs/2102.03152}{arXiv:2102.03152} [gr-qc]].


\bibitem{Hawking:1973}
S.~W.~Hawking and G.~F.~R.~Ellis,
``The Large Scale Structure of Space-Time'',
\doi{10.1017/CBO9780511524646},
(Cambridge University Press, England, 1973).

\bibitem{Martin-Moruno:2017}
P.~Mart\'\i{}n-Moruno and M.~Visser,
``Generalized Rainich conditions, generalized stress-energy conditions, and the Hawking-Ellis classification'',
Class. Quant. Grav. \textbf{34} (2017) no.22, 225014,
\doi{10.1088/1361-6382/aa9039},
[\href{https://arxiv.org/pdf/1707.04172.pdf}{arXiv:1707.04172} [gr-qc]].

\bibitem{Martin-Moruno:2018}
P.~Mart\'{i}n-Moruno and M.~Visser,
``Essential core of the Hawking\textendash{}Ellis types'',
Class. Quant. Grav. \textbf{35} (2018) no.12, 125003,
\doi{10.1088/1361-6382/aac147},
[\href{https://arxiv.org/pdf/1802.00865.pdf}{arXiv:1802.00865} [gr-qc]].

\bibitem{Martin-Moruno:2021}
P.~Mart\'{i}n-Moruno and M.~Visser,
``Hawking-Ellis classification of stress-energy tensors: Test fields versus backreaction'',
Phys. Rev. D \textbf{103} (2021) no.12, 124003,
\doi{10.1103/PhysRevD.103.124003},
[\href{https://arxiv.org/pdf/2102.13551.pdf}{arXiv:2102.13551} [gr-qc]].

\end{thebibliography}
\end{document}